\def\lessthanorabout
  {\mathrel{\raise.3ex\hbox{$<$\kern-.75em\lower1ex\hbox{$\sim$}}}}
\def\greaterthanorabout
  {\mathrel{\raise.3ex\hbox{$>$\kern-.75em\lower1ex\hbox{$\sim$}}}}

\documentstyle[aps,prb,psfig]{revtex}
\begin{document}
\bibliographystyle{jcp}
\title{Quenched degrees of freedom in symmetric diblock copolymer thin films}
\author{Wilfred H. Tang$^*$ \and
        Thomas A. Witten \\
James Franck Institute\\
University of Chicago\\
Chicago, IL 60637\\
}
\maketitle
\begin{abstract}
We study the effect of monomer immobilization (quenching) on the orientation
of the lamellae in symmetric diblock copolymer thin films with neutrally
wetting surfaces.  A small fraction of the monomers
immediately next to the solid substrate is presumed to be quenched.
In both the weak segregation limit and the strong segregation
limit, quenching favors the lamellae orienting perpendicular
to the film.  Quenching inhibits the order--disorder transition twice
as much for the parallel orientation as for the perpendicular.
\end{abstract}

\section{Introduction}
Bulk diblock copolymer melts have been studied extensively, both
experimentally and theoretically.\cite{blockreview1,blockreview2}
Recently, many studies have examined thin films of diblock copolymers.
In this paper, we focus on symmetric diblock copolymers.  In the bulk,
symmetric diblock copolymers form lamellae upon microphase separation.
In thin films, a variety of behavior has been observed.
In some instances, the copolymers form lamellae parallel to the film
interfaces.\cite{flat1,flat2,flat3,flat4}
(See figure 1(a).)  Each of the two interfaces usually
preferentially attracts one of the copolymer species.  Having the lamellae
parallel to the interfaces allows the maximum amount of favorable contact.
But in many experiments, other morphologies are
observed.\cite{henkee,kellogg,koneripalli,morkved,chaikin}
(See, for example, figures 1(b) and 1(c).)
Some of the results can be explained by
confinement.\cite{kikuchibinder,walton,brownchakrabarti,matsen,pickett}
Frustration resulting from the film thickness being
incommensurate with the natural lamellar period can cause the morphology
of figure 1(a) to become energetically unfavorable.
However, confinement does not appear to be sufficient
to explain all the experimental observations.
In this paper, we study the effect of monomer mobility on the morphology.
There is reason to believe that, at least in some situations, the
mobility of some monomers immediately adjacent to a solid substrate is
greatly reduced.\cite{kremer,mobilityexperiment1,mobilityexperiment2}
To make the calculation tractable, we assume that
such monomers are completely immobilized (that is, {\it quenched}).

Our system is a thin polymer film of thickness $h$.
Each polymer has $N/2$ monomers of some species A and $N/2$ monomers
of species B.
The $z=0$ surface of the film is adjacent to a solid substrate, while the
other surface $z=h$ is free (in contact with air, for example).
Since we are interested in the effect of quenching, not confinement, we
assume that the film can adjust its thickness to eliminate any
incompatibility between the film thickness and the natural lamellar period.
We assume that the film thickness $h$ is moderate --- large compared
to the chain root-mean-square end-to-end distance $R_e$ but not so large 
that the bulk properties overpower the interfacial effects.
A small fraction of monomers immediately adjacent to the solid substrate
at $z=0$ is assumed to be quenched.
To simplify the calculation, we assume that the fraction of
quenched monomers is sufficiently small that at most one monomer per
chain is quenched.
We also assume that it is equally likely for A and B monomers to be quenched.
Finally, we assume that both
the solid substrate and the free surface are neutral --- that is, there is
no preferential segregation of A or B monomers to the substrate or the
free surface.  This last assumption differs from the conditions
in most experiments, though through careful design, it is possible to satisfy
this last assumption.\cite{kellogg,neutralmansky1,neutralmansky2}
In future studies, we would like to eliminate this final assumption.
We consider both the weak segregation limit and the strong segregation limit.
Previous theories\cite{fredrickson} have treated the effect of surfaces
on microphase separation, but they have not considered the effect of quenching.
We show that monomer immobilization alone is sufficient to change the phase
transition.
It should be noted that the ability of quenching to alter the properties of
materials in nonobvious ways has been demonstrated in a large variety of
systems.\cite{otherquenching1,otherquenching2,otherquenching3,otherquenching4}

\section{Weak segregation limit}
Our formalism for the weak segregation limit is that of
Marko and Witten\cite{marko,dongmarko}
and is similar in spirit to that of Leibler.\cite{leibler}
First, consider a reference (unquenched)
system in which the A and B monomers are
chemically identical --- that is, each polymer chain consists of $N$
identical monomers, the first half labelled A and the other
half labelled B.  The partition function for this system is
\begin{equation}
Z_{\rm ref} = \int d{\bf R} e^{-S[{\bf R}]}
\end{equation}
where $\int d{\bf R}$ is an integral over monomer positions of all chains
and $S[{\bf R}]$ is the free energy arising from chain connectivity.
The equilibrium average of an arbitrary quantity $X[{\bf R}]$ is given by
\begin{equation}
\langle X \rangle_{\rm ref}
  = {{\int d{\bf R} e^{-S[{\bf R}]} X[{\bf R}]} \over
    {\int d{\bf R} e^{-S[{\bf R}]}}}
\end{equation}
Next, apply a perturbation free energy that arises from the immiscibility
of A and B monomers:
\begin{equation}
S' = - \int d{\bf r} \mu({\bf r}) \rho({\bf r})
\end{equation}
where ${\bf r} = (x,y,z)$ denotes the three spatial coordinates,
$\rho({\bf r}) = \rho_{\rm A}({\bf r}) - \rho_{\rm B}({\bf r})$ is the order
parameter for phase separation,
and $\rho_{\rm A}({\bf r})$ and $\rho_{\rm B}({\bf r})$ are the local
volume fractions of A and B monomers respectively.
(Note that $\rho_{\rm A}({\bf r}) + \rho_{\rm B}({\bf r}) = 1$.)
The partition function for the perturbed system is
\begin{equation}
Z = \int d{\bf R} e^{-S[{\bf R}]}
               e^{S'}
\label{eq:Z}
\end{equation}
and the equilibrium average is given by
\begin{equation}
\langle X \rangle
  = {{\int d{\bf R} e^{-S[{\bf R}]}
    e^{S'} X[{\bf R}]}
    \over
    {\int d{\bf R} e^{-S[{\bf R}]}}
    e^{S'}}
\end{equation}
Since the weak segregation limit corresponds to small perturbations
(small $\mu$), we may express the order parameter as an expansion in the
external potential $\mu$:
\begin{equation}
\langle \rho({\bf r_1}) \rangle
  = {{\partial \ln Z} \over {\partial \mu({\bf r_1})}}
  = G^{(1)}({\bf r_1}) +
    \int d{\bf r_2} G^{(2)}({\bf r_1},{\bf r_2}) \mu({\bf r_2}) +
    {\mathcal{O}}(\mu^2)
\label{eq:linearresponse}
\end{equation}
where $G^{(1)}({\bf r_1})$ and $G^{(2)}({\bf r_1},{\bf r_2})$ are defined as
\begin{equation}
G^{(1)}({\bf r_1}) =
    \left. {{\partial \ln Z} \over {\partial \mu({\bf r_1})}}
    \right| _{\mu = 0}
\end{equation}
and
\begin{equation}
G^{(2)}({\bf r_1},{\bf r_2}) = 
    \left. {{\partial^2 \ln Z} \over {\partial \mu({\bf r_1})
    \partial \mu({\bf r_2})}} \right| _{\mu=0}
\end{equation}
Using equation~\ref{eq:Z}, we find that
\begin{equation}
G^{(1)}({\bf r_1}) =
    \langle \rho({\bf r_1}) \rangle _{\rm ref} = 0
\end{equation}
and
\begin{equation}
G^{(2)}({\bf r_1},{\bf r_2})
  = \langle \rho({\bf r_1}) \rho({\bf r_2}) \rangle _{\rm ref} -
    \langle \rho({\bf r_1}) \rangle _{\rm ref}
    \langle \rho({\bf r_2}) \rangle _{\rm ref}
  = \langle \rho({\bf r_1}) \rho({\bf r_2}) \rangle _{\rm ref}
\label{eq:G2noquench}
\end{equation}
Thus equation~\ref{eq:linearresponse} can be rewritten (neglecting
$\mu^{2}$ and higher order terms) as
\begin{equation}
\langle \rho({\bf r_1}) \rangle  =
    \int d{\bf r_2} G^{(2)}({\bf r_1},{\bf r_2}) \mu({\bf r_2})
\label{eq:linearresponse2}
\end{equation}

In our copolymer melt, we use the Flory-Huggins form for the
free energy perturbation\cite{degennes}
\begin{equation}
S_{\rm FH} = \Lambda \int d{\bf r} \rho_{\rm A}({\bf r})
               \rho_{\rm B}({\bf r})
             = - {\Lambda \over 4} \int d{\bf r} \rho^2({\bf r}) +
               {\rm constant}
\end{equation}
The demixing parameter\cite{roe} $\Lambda$ has units of inverse volume and is
related to the Flory $\chi$ parameter\cite{degennes} by $\Lambda V = \chi N$,
where $V$ is the chain volume and $N$ is the number of monomers per chain.
Using a mean-field approximation,
\begin{equation}
\mu({\bf r}) = - \left\langle {{\partial S_{\rm FH}} \over
                               {\partial \rho({\bf r})}}
               \right\rangle 
             = {\Lambda \over 2} \langle \rho({\bf r}) \rangle 
\end{equation}
and equation~\ref{eq:linearresponse2} becomes an integral eigenvalue equation:
\begin{equation}
\langle \rho({\bf r_1}) \rangle  =
    {\Lambda \over 2} \int d{\bf r_2} G^{(2)}({\bf r_1},{\bf r_2})
    \langle \rho({\bf r_2}) \rangle 
\label{eq:eigenvaluenoquench}
\end{equation}
For small values of $\Lambda$, the only solution to
equation~\ref{eq:eigenvaluenoquench} is
$\langle \rho({\bf r}) \rangle = 0$
everywhere --- that is,
$\langle \rho_{\rm A}({\bf r}) \rangle =
 \langle \rho_{\rm B}({\bf r}) \rangle$
everywhere, and the copolymer melt is structureless.
The smallest value of $\Lambda$ which yields a nontrivial 
$\langle \rho({\bf r}) \rangle$ marks the onset of
microphase separation.

For a bulk copolymer melt, the results are
well known.
$G^{(2)}({\bf r_1},{\bf r_2})
 = 2\langle \rho_{\rm A}({\bf r_1}) \rho_{\rm A}({\bf r_2}) \rangle _{\rm ref}
  -2\langle \rho_{\rm A}({\bf r_1}) \rho_{\rm B}({\bf r_2}) \rangle _{\rm ref}$
is readily calculated from random walk statistics.
There can be no contribution to the $\langle \cdots \rangle$'s unless
${\bf r_1}$ and ${\bf r_2}$ are on the same polymer.
Thus the averages can be calculated by summing over all possible pairs
of monomers along a single chain.  It is convenient to label a monomer
by the volume $v$ displaced by the section of chain between that monomer
and the A end of the chain.  We label the monomer at ${\bf r_1}$ by $v_1$
and the monomer at ${\bf r_2}$ by $v_2$.  Thus
\begin{eqnarray}
G^{(2)}({\bf r_1},{\bf r_2})
 &=&{2 \over V} \int_0^{V/2} dv_1 \int_0^{V/2} dv_2
    \left({a \over {2 \pi \left| v_1 - v_2 \right|}} \right) ^{3/2}
    e^{-{a \over {2 \left| v_1 - v_2 \right|}}
       \left| {\bf r_1} - {\bf r_2} \right| ^2}
\nonumber
\\
 &&-{2 \over V} \int_0^{V/2} dv_1 \int_{V/2}^V dv_2
    \left({a \over {2 \pi \left| v_1 - v_2 \right|}} \right) ^{3/2}
    e^{-{a \over {2 \left| v_1 - v_2 \right|}}
       \left| {\bf r_1} - {\bf r_2} \right| ^2}
\label{eq:G2randomwalk}
\end{eqnarray}
where $a$ is the ``packing length''\cite{fetters} and $V$ is the volume
displaced by an entire chain.
Since in this case $G^{(2)}({\bf r_1},{\bf r_2})$ depends on ${\bf r_1}$
and ${\bf r_2}$ through ${\bf r_1}-{\bf r_2}$ only, the eigenvalue
equation~\ref{eq:eigenvaluenoquench} can be readily solved using
Fourier transforms and the convolution theorem:
\begin{equation}
\langle \rho({\bf k}) \rangle  =
    {\Lambda \over 2} G^{(2)}({\bf k}) \langle \rho({\bf k}) \rangle 
\end{equation}
The onset of microphase separation occurs at
$\Lambda V=10.495$ and $k^*=4.77/R_e$, 
where $R_e$ is the root-mean-square end-to-end distance;
$\langle \rho({\bf r}) \rangle $ can be any sinusoidal function with a
wavevector of magnitude $k^*$.
A convenient basis set for $\langle \rho({\bf r}) \rangle $ consists of
functions of the form
\begin{equation}
\psi_i({\bf r}) =
\left\{ \begin{array}{ll}
            \cos(k_x x) & [k_x>0] \\
            1 & [k_x=0] \\
            \sin(k_x x) & [k_x>0]
        \end{array}
\right\}
\times
\left\{ \begin{array}{ll}
            \cos(k_y y) & [k_y>0] \\
            1 & [k_y=0] \\
            \sin(k_y y) & [k_y>0]
        \end{array}
\right\}
\times
\left\{ \begin{array}{ll}
            \cos(k_z z) & [k_z>0] \\
            1 & [k_z=0] \\
            \sin(k_z z) & [k_z>0]
        \end{array}
\right\}
\label{eq:psibulk}
\end{equation}
where the factors in the three braces are multiplied together in all possible
combinations such that $k_x^2 + k_y^2 + k_z^2$ is satisfied.
(There are 26 total combinations; $ 1 \times 1 \times 1$
does not satisfy $k_x^2 + k_y^2 + k_z^2 = k^{*2}$.)
Thus $\langle \rho({\bf r}) \rangle = \sum_i a_i \psi_i({\bf r})$, where
the $a_i$ are arbitrary constants.
To obtain further information about the microphase separation, we must
consider higher order terms not included in our formalism.
Leibler\cite{leibler} has calculated these higher order terms for all
the possible morphologies consistent with $k_x^2 + k_y^2 + k_z^2 = k^{*2}$
and has found that the lamellar morphology is the most stable.

In the calculation described below, we calculate the effect of quenching
before taking into account Leibler's higher order terms.  This is
reasonable if the perturbation to the microphase separation due to
quenching is larger than the perturbation resulting from the higher order
terms.  We can also apply the perturbations in the opposite order.
That is, we can first consider the higher order terms, which tell us
that the most stable morphology is lamellar.  We can then compare the
effect of quenching on the different lamellar orientations.
Such a calculation would be very similar to to the one described below,
and the final result does not change.

In a thin film, the boundaries at $z=0$ and $z=h$ modify the bulk calculation
slightly.  We use Silberberg's chain-swapping procedure, which leads to
reflective boundary conditions.\cite{silberberg,footnotefredrickson}
Thus each of the two Gaussian terms in equation~\ref{eq:G2randomwalk}
should be replaced by
$\exp \left[-{a \over {2 \left| v_1 - v_2 \right|}}
            \left| {\bf r_1} - {\bf r_2} \right| ^2 \right] +
 \exp \left[-{a \over {2 \left| v_1 - v_2 \right|}}
            \left| {\bf \tilde{r}_1} - {\bf r_2} \right| ^2 \right]$,
where ${\bf \tilde{r}_1}$ is the reflection of ${\bf r_1}$ in the $z=0$ plane.
Also, as noted above, the film thickness $h$ is free to adjust.
With these two assumptions, we find that, as in the bulk melt, 
$\Lambda V=10.495$ and $k^*=4.77/R_e$.
However, the reflective boundary condition at $z=0$ restricts the
basis set functions $\psi_i({\bf r})$ of equation~\ref{eq:psibulk}
to functions that are even with respect to $z=0$;
that is, in the thin film, $\langle \rho({\bf r}) \rangle$
is a linear combination of
\begin{equation}
\psi_i({\bf r}) =
\left\{ \begin{array}{ll}
            \cos(k_x x) & [k_x>0] \\
            1 & [k_x=0] \\
            \sin(k_x x) & [k_x>0]
        \end{array}
\right\}
\times
\left\{ \begin{array}{ll}
            \cos(k_y y) & [k_y>0] \\
            1 & [k_y=0] \\
            \sin(k_y y) & [k_y>0]
        \end{array}
\right\}
\times
\left\{ \begin{array}{ll}
            \cos(k_z z) & [k_z>0] \\
            1 & [k_z=0]
        \end{array}
\right\}
\label{eq:psifilm}
\end{equation}
where $k_x^2 + k_y^2 + k_z^2 = k^{*2}$.
The reflective boundary condition at $z=h$ restricts $\psi_i({\bf r})$
to functions that are even with respect to $z=h$; thus
$k_z h = n \pi$, where $n$ is a nonnegative integer.

The discussion above applies to unquenched systems.
Applying the same formalism to quenched systems results in modifications to
equations~\ref{eq:G2noquench} and \ref{eq:eigenvaluenoquench}:
\begin{equation}
\overline{G_{\rm q}^{(2)}({\bf r_1},{\bf r_2})} = 
    \overline{\langle \rho({\bf r_1}) \rho({\bf r_2}) \rangle _{\rm ref,q}} -
    \overline{\langle \rho({\bf r_1}) \rangle _{\rm ref,q}
              \langle \rho({\bf r_2}) \rangle _{\rm ref,q}}
\label{eq:G2quench}
\end{equation}
\begin{equation}
\overline{\langle \rho({\bf r_1}) \rangle_{\rm q} } =
    {\Lambda \over 2}
    \int d{\bf r_2} \overline{G_{\rm q}^{(2)}({\bf r_1},{\bf r_2})}
    \overline{\langle \rho({\bf r_2}) \rangle_{\rm q} }
\label{eq:eigenvaluequench}
\end{equation}
where $\langle \cdots \rangle_{\rm ref,q}$ and
$\langle \cdots \rangle_{\rm q}$ denote equilibrium averages subject
to the constraint q and the overbar denotes averaging over the constraints.
In our system, the constraint is that a small fraction of the monomers
immediately adjacent to the solid substrate at $z=0$ cannot move.

The first term of $\overline{G_{\rm q}^{(2)}({\bf r_1},{\bf r_2})}$ in
equation~\ref{eq:G2quench} is equal to the first term of
$G^{(2)}({\bf r_1},{\bf r_2})$ in equation~\ref{eq:G2noquench}
since for any arbitrary quantity $X[{\bf R}]$, 
$\langle X \rangle_{\rm ref} = \overline{\langle X \rangle_{\rm ref,q}}$.
However, while the second, residual term of
$G^{(2)}({\bf r_1},{\bf r_2})$ is zero, the residual term of
$\overline{G_{\rm q}^{(2)}({\bf r_1},{\bf r_2})}$ is nonzero.
Using random walk statistics, we find that for a thin film with quenched
monomers this residual term is
\begin{equation}
\overline{\langle \rho({\bf r_1}) \rangle _{\rm ref,q}
          \langle \rho({\bf r_2}) \rangle _{\rm ref,q}}
  = 2 \overline{\langle \rho_{\rm A}({\bf r_1}) \rangle _{\rm ref,q}
                \langle \rho_{\rm A}({\bf r_2}) \rangle _{\rm ref,q}} -
    2 \overline{\langle \rho_{\rm A}({\bf r_1}) \rangle _{\rm ref,q}
                \langle \rho_{\rm B}({\bf r_2}) \rangle _{\rm ref,q}}
\label{eq:residualtermbase}
\end{equation}
where
\begin{eqnarray}
\lefteqn{\overline{\langle \rho_{\rm A}({\bf r_1}) \rangle _{\rm ref,q}
                   \langle \rho_{\rm A}({\bf r_2}) \rangle _{\rm ref,q}}}
\nonumber
\\
 &=&4 \sigma \int dx' \int dy'
    {1 \over V} \int_0^V dv
    \int_0^{V/2} dv_1
    \left( {a \over {2 \pi \left| v-v_1 \right| }} \right) ^{3/2}
    e^{-{a \over {2 \left| v-v_1 \right|}} [(x_1-x')^2+(y_1-y')^2+z_1^2]}
\nonumber
\\
  &&\int_0^{V/2} dv_2
    \left( {a \over {2 \pi \left| v-v_2 \right| }} \right) ^{3/2}
    e^{-{a \over {2 \left| v-v_2 \right|}} [(x_2-x')^2+(y_2-y')^2+z_2^2]}
\nonumber
\\
 &=&{{4 \sigma} \over V} \int_0^V dv \int_0^{V/2} dv_1 \int_0^{V/2} dv_2
    {a \over {2 \pi ( \left| v-v_1 \right| + \left| v-v_2 \right| )}}
    e^{-{a \over {2 ( \left| v-v_1 \right| + \left| v-v_2 \right| )}}
       [(x_1-x_2)^2+(y_1-y_2)^2]}
\nonumber
\\
  &&\left( {a \over {2 \pi \left| v-v_1 \right|}} \right) ^{1/2}
    e^{-{a \over {2 \left| v-v_1 \right|}}z_1^2}
    \left( {a \over {2 \pi \left| v-v_2 \right|}} \right) ^{1/2}
    e^{-{a \over {2 \left| v-v_2 \right|}}z_2^2}
\label{eq:residualtermpartone}
\end{eqnarray}
and
$\overline{\langle \rho_{\rm A}({\bf r_1}) \rangle _{\rm ref,q}
           \langle \rho_{\rm B}({\bf r_2}) \rangle _{\rm ref,q}}$
is the same except that the integral of $v_2$ goes from $V/2$ to $V$
instead of $0$ to $V/2$.
The quantity $\sigma$ is the number of quenched monomers per unit area at the
$z=0$ boundary.
As noted above, we assume that the amount of quenching is sufficiently
small that at most one monomer per chain is quenched.  The integral
$\sigma \int dx' \int dy'$ accounts for the fact that it is equally likely
for any boundary location $(x',y',0)$ to have a quenched monomer.
The integral ${1 \over V}\int_0^V dv$ accounts for the fact that it is
equally likely for any monomer of each chain to be quenched.

Since the quenching density $\sigma$ is small,
$\overline{\langle \rho({\bf r_1}) \rangle _{\rm ref,q}
           \langle \rho({\bf r_2}) \rangle _{\rm ref,q}} \ll
 \overline{\langle \rho({\bf r_1}) \rho({\bf r_2}) \rangle _{\rm ref,q}}$,
and we can solve the eigenvalue equation~\ref{eq:eigenvaluequench} using
degenerate perturbation theory.\cite{couranthilbert}
We rewrite equation~\ref{eq:eigenvaluequench} as
\begin{equation}
\int d{\bf r_2}
    \left[
    \overline{\langle \rho({\bf r_1}) \rho({\bf r_2}) \rangle _{\rm ref,q}} -
    \overline{\langle \rho({\bf r_1}) \rangle _{\rm ref,q}
              \langle \rho({\bf r_2}) \rangle _{\rm ref,q}}
    \right]
    \overline{\langle \rho({\bf r_2}) \rangle_{\rm q} } =
    \lambda \overline{\langle \rho({\bf r_1}) \rangle_{\rm q} }
\label{eq:eigenvaluequench2}
\end{equation}
where $\lambda = 2/\Lambda$.
In the absence of quenching, the solution to
equation~\ref{eq:eigenvaluequench2} is 
$\lambda^{(0)} = 2 / \Lambda^*$ and
$\overline{\langle \rho({\bf r}) \rangle_{\rm q} }^{(0)}
 = \sum_i a_i \psi_i({\bf r})$, where the $\psi_i({\bf r})$ of
equation~\ref{eq:psifilm} form a basis set for the unperturbed solutions
and the $a_i$ are arbitrary constants.
In the presence of a small amount of quenching, we can perform an
expansion in $\sigma$ about
$\lambda^{(0)}$ and
$\overline{\langle \rho({\bf r}) \rangle_{\rm q} }^{(0)}$:
\begin{equation}
\lambda = \lambda^{(0)} + \sigma \lambda^{(1)} + \sigma^2 \lambda^{(2)}
          + \cdots
\end{equation}
\begin{equation}
\overline{\langle \rho({\bf r}) \rangle_{\rm q} } =
    \overline{\langle \rho({\bf r}) \rangle_{\rm q} }^{(0)} +
    \sigma \overline{\langle \rho({\bf r}) \rangle_{\rm q} }^{(1)} +
    \sigma^2 \overline{\langle \rho({\bf r}) \rangle_{\rm q} }^{(2)} +
    \cdots
\end{equation}
Applying degenerate perturbation theory and requiring that
$\int d{\bf r} \psi_i({\bf r}) \psi_j({\bf r}) = 0$ for $i \neq j$,
we can calculate $\sigma \lambda^{(1)}$ by solving
\begin{equation}
-\sum_i a_i \int d{\bf r_1} \int d{\bf r_2}
    \overline{\langle \rho({\bf r_1}) \rangle _{\rm ref,q}
              \langle \rho({\bf r_2}) \rangle _{\rm ref,q}}
    \psi_j({\bf r_1}) \psi_i({\bf r_2}) -
    \sigma \lambda^{(1)} a_j \int d{\bf r} \psi_j^2({\bf r}) = 0
\label{eq:lambda1}
\end{equation}
From equations~\ref{eq:psifilm}, \ref{eq:residualtermbase},
and \ref{eq:residualtermpartone}, we see that
$\int d{\bf r_1} d{\bf r_2}
    \overline{\langle \rho({\bf r_1}) \rangle _{\rm ref,q}
              \langle \rho({\bf r_2}) \rangle _{\rm ref,q}}
    \psi_j({\bf r_1}) \psi_i({\bf r_2}) = 0$ for $i \neq j$.\cite{footnoteij}
Thus our original eigenfunctions $\psi_i$ are already adequate to
treat the perturbed system, and
equation~\ref{eq:lambda1} reduces to
\begin{equation}
\sigma \lambda^{(1)} =
  -{{\int d{\bf r_1} \int d{\bf r_2}
    \overline{\langle \rho({\bf r_1}) \rangle _{\rm ref,q}
              \langle \rho({\bf r_2}) \rangle _{\rm ref,q}}
    \psi_i({\bf r_1}) \psi_i({\bf r_2})} \over
   {\int d{\bf r} \psi_i^2({\bf r})}}
\end{equation}
The integrals $\int d{\bf r}$, $\int d{\bf r_1}$, and $\int d{\bf r_2}$
range over the volume of the thin film.  We assume that the extent of the
film in the $x$ and $y$ directions is large enough that the $x$ and $y$
boundaries have a negligible effect.  In the $z$ direction, the integrals
run from $0$ to $h$.  As noted above, $h$ is assumed to be
large compared to the polymer rms end-to-end distance $R_e$.  Thus it is a good
approximation to replace $\int_0^h dz_1$ and $\int_0^h dz_2$ in the
numerator by $\int_0^\infty dz_1$ and $\int_0^\infty dz_2$.

For eigenfunctions with nonzero $k_z$,
\begin{equation}
\psi_i({\bf r}) =
\left\{ \begin{array}{ll}
            \cos(k_x x) & [k_x>0] \\
            1 & [k_x=0] \\
            \sin(k_x x) & [k_x>0]
        \end{array}
\right\}
\times
\left\{ \begin{array}{ll}
            \cos(k_y y) & [k_y>0] \\
            1 & [k_y=0] \\
            \sin(k_y y) & [k_y>0]
        \end{array}
\right\}
\times
        \begin{array}{ll}
            \cos(k_z z) & [k_z>0]
        \end{array}
\label{eq:psibig}
\end{equation}
where $k_x^2+k_y^2+k_z^2 = k^{*2}$, we find that
\begin{equation}
\sigma \lambda^{(1)} = - {{4 c \sigma V^2} \over h}
\end{equation}
and
\begin{equation}
\Lambda
  = {2 \over \lambda}
  \approx {2 \over {\lambda^{(0)} + \sigma \lambda^{(1)}}}
  \approx \Lambda^{(0)}
          \left( 1-{\Lambda^{(0)} \over 2} \sigma \lambda^{(1)} \right)
  = \Lambda^{(0)} \left( 1+{{2 c \sigma V^2 \Lambda^{(0)}} \over h} \right)
\label{eq:Lambdabig}
\end{equation}
where $\Lambda^{(0)}$ corresponds to the microphase separation transition
in the absence of quenching
and $c$ is a constant defined as\cite{footnotec}
\begin{eqnarray}
c &\equiv&
  {1 \over {V^3}} \int_0^V dv \int_0^{V/2} dv_1 \int_0^{V/2} dv_2
  e^{-{ {\left| v-v_1 \right| + \left| v-v_2 \right|} \over {2 a}} k^{*2}}
\nonumber
\\
  &&-
  {1 \over {V^3}} \int_0^V dv \int_0^{V/2} dv_1 \int_{V/2}^V dv_2
  e^{-{ {\left| v-v_1 \right| + \left| v-v_2 \right|} \over {2 a}} k^{*2}}
\nonumber
\\
  &\approx& 0.0204
\end{eqnarray}
On the other hand, for eigenfunctions with $k_z=0$,
\begin{equation}
\psi_i({\bf r}) =
\left\{ \begin{array}{ll}
            \cos(k_x x) & [k_x>0] \\
            1 & [k_x=0] \\
            \sin(k_x x) & [k_x>0]
        \end{array}
\right\}
\times
\left\{ \begin{array}{ll}
            \cos(k_y y) & [k_y>0] \\
            1 & [k_y=0] \\
            \sin(k_y y) & [k_y>0]
        \end{array}
\right\}
\times
        \begin{array}{ll}
            1 & [k_z=0]
        \end{array}
\label{eq:psismall}
\end{equation}
where $k_x^2+k_y^2+k_z^2 = k^{*2}$, we find that
\begin{equation}
\sigma \lambda^{(1)} = - {{2 c \sigma V^2} \over h}
\end{equation}
and
\begin{equation}
\Lambda \approx
  \Lambda^{(0)} \left( 1+{{c \sigma V^2 \Lambda^{(0)}} \over h} \right)
\label{eq:Lambdasmall}
\end{equation}

We are interested in the eigenfunctions with the smallest value of $\Lambda$.
Thus, in the quenched system, the microphase separation
transition occurs at $\Lambda$ given by equation~\ref{eq:Lambdasmall} and
$\langle \rho({\bf r}) \rangle = \sum_i a_i \psi_i({\bf r})$, where
the $\psi_i({\bf r})$ are given by equation~\ref{eq:psismall} and the
$a_i$ are constants.
Taking into account the higher order terms calculated by Leibler\cite{leibler}
further restricts $\langle \rho({\bf r}) \rangle$ to lamellar morphologies.
The only lamellar $\langle \rho({\bf r}) \rangle$ that can be constructed using
the $\psi_i({\bf r})$ of equation~\ref{eq:psismall} as basis functions are
$\langle \rho({\bf r}) \rangle = \cos(k_x x + k_y y + \theta)$ 
where $k_x^2+k_y^2 = k^{*2}$ and $\theta$ is a phase constant ---
that is, the lamellae are perpendicular to the substrate, as in
figure 1(b).  (Notice that having lamellae parallel to the substrate ---
that is, $\langle \rho({\bf r}) \rangle = \cos(k^* z)$ ---
would require $\psi_i({\bf r})$ given by equation~\ref{eq:psibig},
corresponding to $\Lambda$ given by equation~\ref{eq:Lambdabig}.  Since this
is not the minimum value of $\Lambda$, the $\psi_i({\bf r})$ given by
equation~\ref{eq:psibig} cannot be used.)

Thus quenching inhibits phase separation since quenching increases the
value of the demixing parameter $\Lambda$ at the order--disorder transition.
Furthermore, quenching inhibits the formation of lamellae parallel to the
substrate twice as much as lamellae perpendicular to the substrate.
Consequently, quenching favors the lamellae orienting perpendicular to
the substrate.  From equation~\ref{eq:Lambdasmall}, we see that the
amount of inhibition increases with increasing quenching density $\sigma$
and decreases with increasing film thickness $h$.
Since quenching is an interfacial effect,
increasing the volume of material relative to the interfacial area
decreases the effect of quenching.  We can also estimate the order of magnitude
of the increase in $\Lambda$ due to quenching.  A large change in $\Lambda$
would require large $\sigma$ and small $h$.  The largest monomer quenching
density consistent with the assumptions in our calculation is
$\sigma \approx a/V$, corresponding to about one quenched monomer per chain
touching the substrate.  Our formalism can handle higher quenching
densities, but equation~\ref{eq:residualtermpartone} would need to be modified
to allow for more than one quenched monomer per chain.
The smallest film thickness allowed by our calculation is
$h \approx R_e \approx \sqrt{V/a}$.
Having $h$ smaller than $R_e$ would cause our calculation to break down
in several places.  Thus the largest possible perturbation consistent
with our assumptions is
\begin{equation}
{{\Delta \Lambda} \over {\Lambda^{(0)}}}
  = {{c \sigma V^2 \Lambda^{(0)}} \over h}
  \approx c {{a^{3/2}} \over {V^{1/2}}} \Lambda^{(0)} V
\end{equation}
For a typical polymer, the chain volume $V \approx 10^5 {\rm \AA}$ and the
packing length $a \approx 10 {\rm \AA}$.
We also know that $c \approx 0.02$ and $\Lambda^{(0)} V \approx 10$, so
$\Delta \Lambda / \Lambda^{(0)} \approx 0.02$.
A 2\% change in the demixing parameter $\Lambda$ could be difficult to
detect experimentally but is not so small that one can ignore it completely
in interpreting experimental results.

\section{Strong segregation limit}
In the strong segregation limit, we assume that symmetric diblock copolymers
form lamellae consisting of regions of A monomers separated by sharp
interfaces from regions of B monomers.
In a thin film, the lamellae can orient either parallel or perpendicular
to the substrate.
In the absence of quenching, the free energies of these two orientations are
equal, but this degeneracy is broken when there is quenching.
Consider the effect of quenched monomers when the lamellae
are parallel to the substrate, as in figure 1(a).  Half of the
quenched monomers are stuck in an unfavorable region.  In order for a
polymer chain containing an unfavorably quenched monomer to reach a favorable
region, the disfavored chain segment must stretch a distance $L/4$, where
$L$ is the lamellar period, and this costs substantial free energy.
Now consider the effect of quenched monomers when the lamellae are
perpendicular to the substrate, as in figure 1(b).  Again, half
of the quenched monomers are stuck in an unfavorable region.  In
order for a polymer chain containing an unfavorably quenched monomer to reach
a favorable region, the disfavored chain segment must stretch a distance of
{\it at most} $L/4$.  Most chains containing an unfavorably quenched monomer
need only stretch a shorter distance and thus incur a smaller free energy
penalty.  Therefore, as in the weak segregation limit, quenching
causes the lamellae to orient perpendicular to the substrate.

\section{Discussion}
The calculation reported above describes the
effect of immobilized monomers on block copolymer phase separation.
We have shown that this immobilization tends to favor perpendicular lamellae.
We have also shown how the immobilization constraint can be incorporated
into the standard field theory used to treat polymer phase separation.
On the other hand, our theory does not yet give an adequate account of
why perpendicular lamellae are observed experimentally.
Below we discuss the various limitations in our theory,
their experimental relevance, and the prospects for improving the theory.  

The most unrealistic assumption is that there is no preferential
interaction of the A or B monomers with the surface.
Such neutral surfaces are uncommon in practice.
The reports of perpendicular lamellae that motivated this
study\cite{henkee,koneripalli,morkved} were done on surfaces
that were not at all neutral.
Preferential interaction with the surface naturally favors a morphology
in which the surface is covered with the preferred species.
It thus opposes the tendency of the quenched chains to make perpendicular
lamellae.
The preferential surface interaction amounts to a perturbation in the
free energy that is linear in the order parameter $\rho$.
The perturbation due to quenching, on the other hand, is an effect
quadratic in $\rho$:  it does not break the A--B symmetry.
Thus, for any significant phase separation amplitude, we expect the
linear preferential adsorption term to dominate, unless its coefficient
is zero (i.e., neutral surfaces).

In any case, even if the surfaces are nearly neutral, there are no lack of
effects favoring perpendicular orientation, even without quenching.
The effect of incommensurate
thickness\cite{kikuchibinder,walton,brownchakrabarti,matsen,pickett}
was already mentioned in the introduction.
In addition, nematic interactions between the monomers and the surface
favor perpendicular lamellae.\cite{nematic}
Given all these restrictions, we do not expect quenching to play an
important role in the weak segregation regime.  In the strong segregation
regime, the quenched chains carry a large free energy penalty and could
play a much greater role.  Our analysis of this regime is in progress.

Another restriction of our theory is the assumption of a small number of
immobilized monomers.  Under this assumption the effect on the phase
separation threshold $\Lambda$ is of course small as well.
Still, the calculation can give a glimpse of the effects of larger quenching
density.  If the quenched monomer density approaches one per chain touching
the surface, the volume fraction of quenched chains near the surface
becomes of order unity.  Beyond this point our calculation becomes unreliable.
As the number of immobilized monomers increases beyond this point,
the surface chains become quenched at multiple sites, thus forming a
series of grafted loops.  One anticipates that these loops would
inhibit microphase separation even more than the singly-attached chains do,
since the fluctuations that allow phase separation are more inhibited.

Our results are amenable to quantitative tests.
Such tests would be of interest, as they would reveal how immobilization
affects phase separation morphology.  Naturally, one possible test is to
induce block copolymer phase separation on a neutral surface containing a
few grafted chains.  Though we have only treated chains immobilized at an
arbitrary monomer, our method is easily adapted to treat various cases ---
for example, end-grafted chains or chains grafted at the A--B junction point.
One need only replace the average ${1 \over V}\int_0^V dv$ in
equation~\ref{eq:residualtermpartone} by the appropriate superposition
of $v$ values.  An experimental test could reveal the optimal way of
influencing the morphology.  It could also extend the understanding of
the quenching effect beyond the narrow limits treated above.
Such an experimental test, however, would face obstacles.
Any residual non-neutrality of the surfaces would inhibit perpendicular
orientation.  In addition, nematic interactions with the surface would favor
perpendicular orientation even in the absence of quenching, as noted above.
A cleaner investigation could be performed using computer simulations.
Here the neutral surface and the A-B symmetry of the chains could be
realized exactly.  Such simulations could help elucidate the interplay
between the various factors, including preferential surface interactions,
nematic interactions, film confinement, and quenching, that influence
the phase separation morphology.

\section{Conclusions}
We have shown that monomer mobility can affect the morphology of diblock
copolymer thin films.  In our system, immobilizing a fraction of the monomers
next to the substrate favors the lamellae orienting perpendicular to
the substrate.  In the future we would like to study the effect of quenching
in other systems since we believe that quenching occurs to some extent in
a wide variety of systems.

\section*{Acknowledgements}
We thank Terry Morkved, Heinrich Jaeger, and Paul Mansky for helpful
discussions.  
This work was supported in part by the MRSEC Program of the National Science 
Foundation under Award Number DMR-9400379.
W. T. acknowledges support from the Department of Defense through an
NDSEG Fellowship.

\bibliography{quench}

\begin{thebibliography}{10}
\bibitem{footnotefredrickson}
Our reflective boundary conditions are consistent with Fredrickson's boundary
  conditions in reference 22. His treatment allows for a demixing parameter
  $\Lambda$ that is perturbed at the surface. We consider this perturbation to
  be negligible in our calculation.
\bibitem{footnotec}
While c is small, it is not unreasonably small since $1/c \approx 49$ is of the
  same order of magnitude as other constants in our problem, such as
  $\Lambda^{(0)} V \approx 10$ and $k^{*2} R_e^2 \approx 23$.
\end{thebibliography}

\section*{Figure captions}
Figure 1.  Some possible morphologies for thin films of
symmetric diblock copolymers, as suggested by experiments.

\newpage
\begin{figure}
\psfig{figure=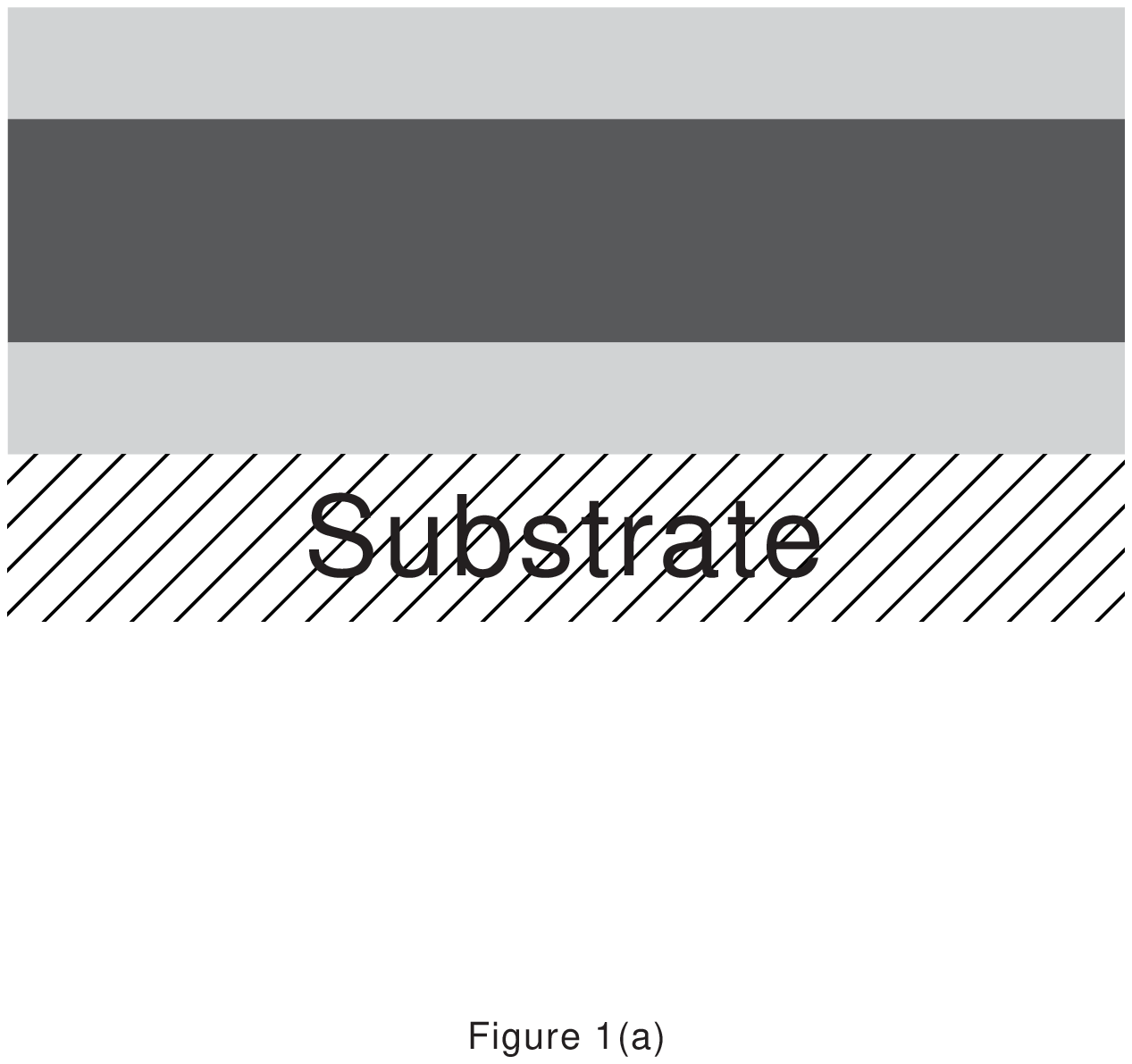}
\end{figure}

\begin{figure}
\psfig{figure=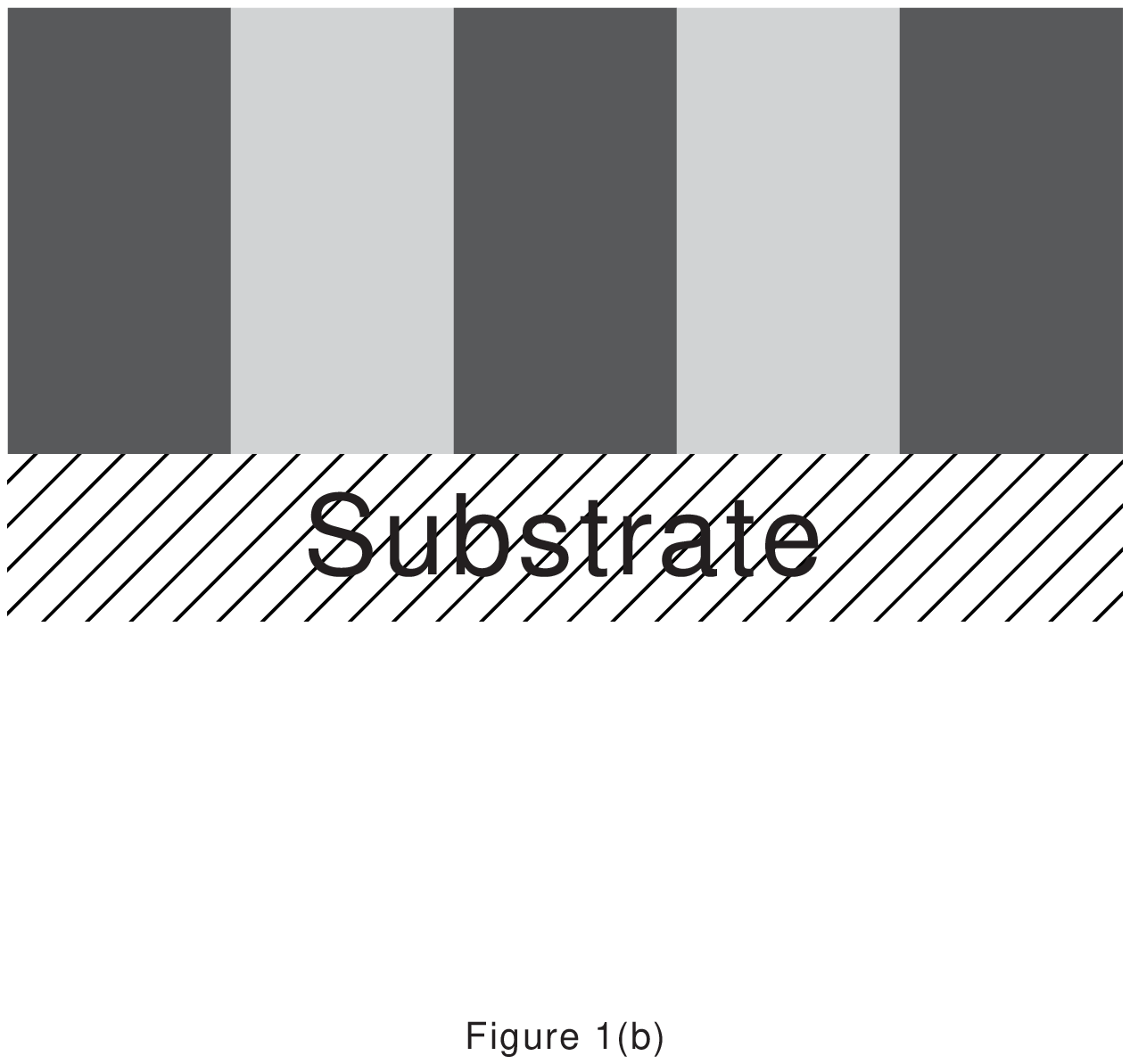}
\end{figure}

\begin{figure}
\psfig{figure=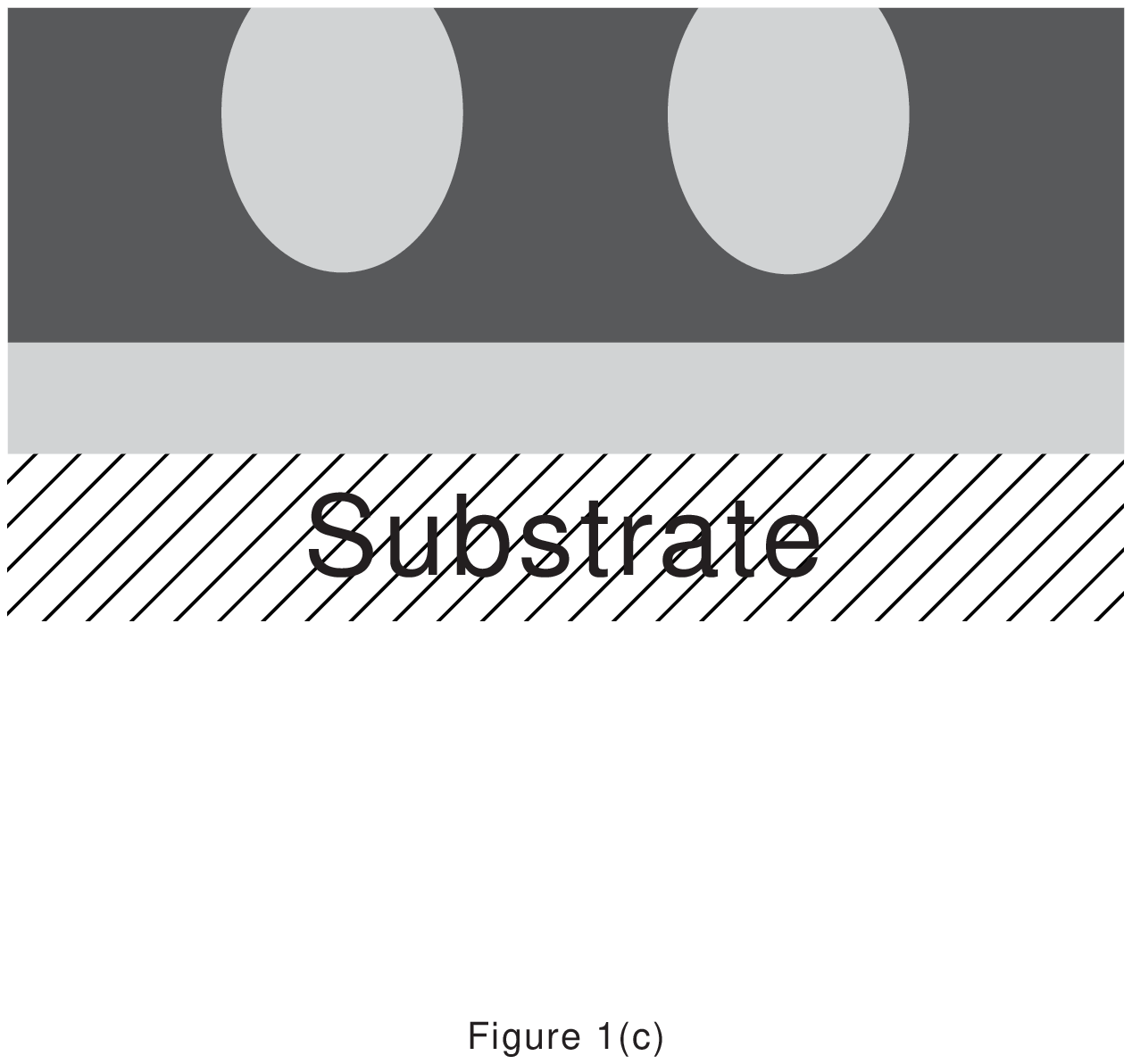}
\end{figure}

\end{document}